\documentclass[aps,prb,twocolumn,superscriptaddress, amsmath,amssymb]{revtex4-2}

\usepackage{soul}
\usepackage{diagbox}
\usepackage{ragged2e}
\usepackage{array}
\usepackage{booktabs}
\usepackage{graphicx}
\usepackage{dcolumn}
\usepackage{bm}
\usepackage[colorlinks, linkcolor=black, anchorcolor=black, citecolor=black]{hyperref}

\begin{document}
\newcommand {\xw}[1]{{\color{magenta} #1}}
\preprint{APS/123-QED}

\title{First-order transitions in weak Ising spin-orbit-coupled superconductors}

\author{Xusheng Wang}
\email{wang-xs23@mails.tsinghua.edu.cn}
\affiliation{%
State Key Laboratory of Low-Dimensional Quantum Physics, Department of Physics, Tsinghua University, Beijing 100084, China}
\author{Gaomin Tang}
\affiliation{
Graduate School of China Academy of Engineering Physics, Beijing 100193, China}
\author{Shuai-Hua Ji}
\affiliation{%
State Key Laboratory of Low-Dimensional Quantum Physics, Department of Physics, Tsinghua University, Beijing 100084, China}
\affiliation{%
Frontier Science Center for Quantum Information, Beijing 100084, China}


\begin{abstract}

Ising spin-orbit coupling (ISOC) can strongly protect superconductivity against exchange-field-induced depairing, typically leading to critical fields far exceeding the Pauli limit and continuous (second-order) phase transitions. Here, using a standard free-energy approach, we demonstrate that first-order transitions can emerge in superconductors with weak ISOC under large exchange fields. In this regime, theoretical approaches based on the gap equation fail to determine the thermodynamic critical field and instead yield only the supercooling field. We further identify two pronounced in-gap coherence peaks in the quasiparticle spectra, which represent the weak-ISOC manifestation of the previously reported mirage-gap states. These thermodynamic and spectroscopic signatures remain robust against modest impurity scattering and provide experimentally accessible criteria for identifying weak Ising superconductivity, which serves as an intermediate regime between conventional Pauli-limited and strong Ising superconductors.
\end{abstract}

\maketitle

\section{Introduction}

For two-dimensional superconductors, in-plane magnetic fields effectively act as exchange fields, providing an ideal platform to study the competition between superconductivity and Zeeman effect. Conventional superconductivity, characterized by spin-singlet pairing, is suppressed by a sufficiently large exchange field due to the alignment of electron spins within Cooper pairs. The critical field of isotropic single-band superconductors at zero temperature was first established by Clogston and Chandrasekhar \cite{clogston1962upper,chandrasekhar1962note}, commonly referred to as the Pauli limit, $H_p = 0.707\Delta_0$. Sarma subsequently analyzed the superconducting gap equation at finite exchange field and temperature, revealing multiple solution branches in the low-temperature regime \cite{sarma1963influence}. The corresponding free-energy landscape and the evolution from a low-temperature first-order transition (FOT) to a high-temperature second-order transition (SOT) were elucidated by Maki and Tsuneto \cite{maki1964pauli,maki1966effect}. The Werthamer-Helfand-Hohenberg and Klemm-Luther-Beasley theories further incorporated orbital depairing, impurity and spin-orbit scattering, and dimensional effects, providing the standard framework for assessing how realistic scattering and orbital effects modify or suppress the FOT \cite{werthamer1966temperature,klemm1975theory}. These Pauli-limited predictions and their disorder-dependent evolution were subsequently examined through spin-polarized tunneling and critical-field measurements in ultrathin aluminum films by Tedrow, Meservey, and co-workers \cite{tedrow1970experimental,meservey1970magnetic,meservey1975tunneling,tedrow1977supercooling,tedrow1977measurement}.

Recently, critical fields far exceeding the Pauli limit have been reported in several transition-metal dichalcogenide monolayers \cite{lu2015evidence,saito2016superconductivity,xi2016ising,sohn2018unusual,de2018tuning,cho2022nodal,li2021recent,wan2023orbital,ji2024continuous,volavka2026ising,wickramaratne2020ising}. This enhancement is primarily attributed to Ising spin-orbit coupling (ISOC), which arises from in-plane inversion-symmetry breaking and produces opposite out-of-plane spin polarizations at the $\mathbf{K}$ and $-\mathbf{K}$ valleys \cite{zhu2011giant,xiao2012coupled}. The resulting spin-valley locking suppresses paramagnetic pair breaking by an in-plane magnetic field, thereby strongly enhancing the in-plane critical field \cite{yuan2014possible,zhou2016ising,shaffer2020crystalline}. Previous theoretical studies \cite{yuan2016ising,liu2017unconventional,ilic2017enhancement,liu2020microscopic,mockli2020ising,wang2021ising,ilic2023spectral} have generally assumed that the transition from the superconducting to the normal state in Ising superconductors is continuous (second order). Under this assumption, the critical exchange field can be obtained from the self-consistent gap equation by setting $\Delta = 0$. However, for FOTs, which may arise in superconductors with weak ISOC, this approach yields only the supercooling field \cite{maki1964pauli,fulde1964superconductivity,matsuda2007fulde,olde2021tunable}, rather than the true thermodynamic critical field. In such cases, a full free-energy analysis is required.

In this work, we theoretically investigate the temperature- and exchange-field-dependent superconducting properties of weak Ising superconductors within a standard free-energy framework. We find that SOT dominate both the low- and high-temperature regimes, while an FOT emerges at intermediate temperatures. We further predict two pronounced in-gap coherence peaks in the quasiparticle spectra of weak Ising superconductors. Together, these thermodynamic and spectroscopic features distinguish the weak-ISOC regime from both conventional Pauli-limited superconductors \cite{tedrow1970experimental,tedrow1977measurement} and strong Ising superconductors \cite{ilic2017enhancement,tang2021magnetic}.

\begin{figure*}[htbp]
\centering
\includegraphics[width=0.95\linewidth]{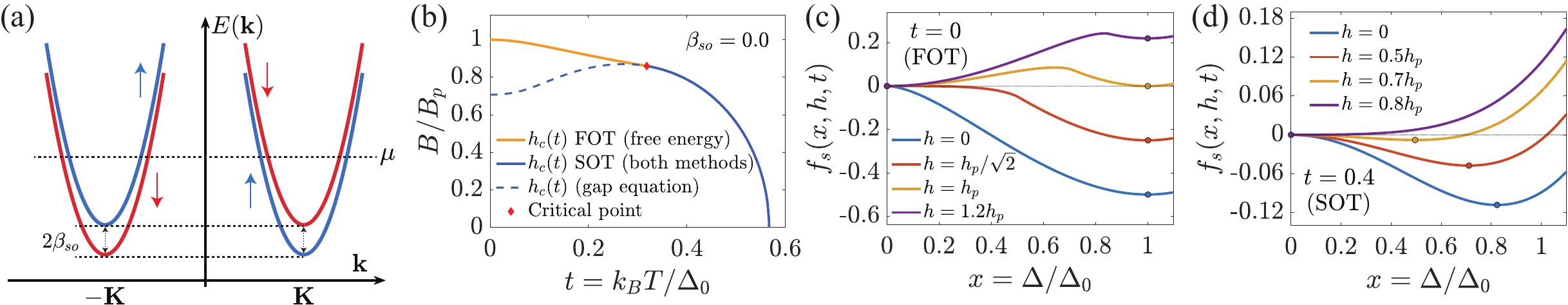}
\caption{
(a) Schematic band structure of the normal state with ISOC. $\mathbf{K}$ and $-\mathbf{K}$ denote the two valleys, and $\mu$ is the chemical potential.
(b) Phase diagram in the zero-ISOC limit. The solid yellow curve denotes the FOT critical field obtained from free-energy analysis. The solid blue curve represents the SOT critical field, which can be determined by both the free-energy and gap-equation approaches. The dashed blue curve indicates the supercooling field in the FOT regime, obtained from the gap equation. The red filled diamond marks the critical point separating the FOT and SOT regimes. $B_p$ denotes the Pauli-limit magnetic field.
Free-energy landscape at (c) $t=0$ and (d) $t = 0.4$. In (c) and (d), local minima of free energy are indicated by filled circles, and $h_p$ denotes the dimensionless Pauli-limit field.
}
\label{Ising_illustration}
\end{figure*}

\section{Model and Theoretical Framework}

Considering an $s$-wave spin-singlet superconducting system with ISOC ($\beta_{\mathrm{SO}}$) under an in-plane magnetic field applied along the $x$ axis, the effective normal-state Hamiltonian [see Fig.~\ref{Ising_illustration}(a)] can be written as \cite{ilic2017enhancement}:
\begin{equation}
    \hat{H}_N(\mathbf{k}=\mathbf{p}+\eta\mathbf{K})=\xi_\mathbf{p}\hat{\sigma}_0+\eta \beta_{SO}\hat{\sigma}_z - H \hat{\sigma}_x,
\end{equation}
\noindent where $\mathbf{p}$ is the momentum measured relative to the valley center, $\eta=\pm 1$ labels the two valleys, and $\xi_\mathbf{p}$ denotes the kinetic energy. Here, $\hat{\sigma}_i$ ($i=x,y,z$) are Pauli matrices and $\hat{\sigma}_0$ is the identity matrix in spin space. The Zeeman energy is defined as $H = g_L \mu_B B / 2$, where $g_L$ is the Landé $g$ factor, $\mu_B$ is the Bohr magneton, and $B$ is the applied in-plane magnetic field.

The Bogoliubov-de Gennes Hamiltonian in the Nambu basis $(\hat{c}_{\mathbf{k},\uparrow}, \hat{c}_{\mathbf{k},\downarrow}, \hat{c}_{-\mathbf{k},\uparrow}^\dagger, \hat{c}_{-\mathbf{k},\downarrow}^\dagger)^{\mathrm{T}}$ is given by \cite{tang2021magnetic,patil2023spectral,harms2025collapse}:
\begin{equation}
    \hat{H}_{\mathrm{BdG}}(\mathbf{k})=    
    \begin{pmatrix}
    \hat{H}_N(\mathbf{k}) & i\Delta \hat{\sigma}_y \\
    -i\Delta \hat{\sigma}_y & -\hat{H}^{*}_N(-\mathbf{k})
    \end{pmatrix}.
\end{equation}
\noindent Here, $\Delta$ denotes the isotropic superconducting gap. The positive eigenvalues yield the quasiparticle dispersion:
\begin{equation}\label{quasiparticle}
    E_{\mathbf{p}\pm}=\sqrt{H_{\mathrm{eff}}^2+\Delta^2+\xi_\mathbf{p}^2 \pm 2\sqrt{H^2\Delta^2+H_{\mathrm{eff}}^2\xi_\mathbf{p}^2}},
\end{equation}
\noindent where $H_{\mathrm{eff}}=\sqrt{H^2+\beta_{SO}^2}$ is the effective magnetic field. The free energy of the superconducting state can be expressed as \cite{altland2010condensed,xie2020spin,wang2025unified,wang2025temperature}:
\begin{align} \label{free_energy}
    F=\frac{\Delta^2}{g} & + 2N\int_0^{E_c} \left(\xi_\mathbf{p}-\frac{E_{\mathbf{p}+}+E_{\mathbf{p}-}}{2}\right) \mathrm{d}\xi_\mathbf{p} \notag \\
    & - \frac{2N}{\beta}\sum_{\sigma=\pm}\int_0^{E_c}\ln\left(1+e^{-\beta E_{\mathbf{p}\sigma}}\right)\mathrm{d}\xi_\mathbf{p}
\end{align}
\noindent where $g$ is the pairing interaction strength, $N$ is the density of states (DOS) at the Fermi level, $E_c$ is the energy cutoff, and $\beta = 1/(k_B T)$ is the inverse temperature.

Introducing the dimensionless parameters
\begin{equation}
x=\frac{\Delta}{\Delta_0},\quad
h=\frac{H}{\Delta_0},\quad
t=\frac{k_B T}{\Delta_0},\quad
\beta_{\mathrm{so}}=\frac{\beta_{\mathrm{SO}}}{\Delta_0},
\end{equation}

\noindent where $\Delta_0$ is the superconducting gap at zero temperature and zero field, the free energy in Eq.~(\ref{free_energy}) can be recast into a reduced form $f_s = [F(\Delta) - F(0)] / (N\Delta_0^2)$, which takes the normal-state free energy as the reference. The explicit expression is provided in the Supplemental Material \cite{supp}. The physical superconducting gap is determined by minimizing the free energy. Notably, the self-consistent gap equation follows from the stationary condition $\partial f_s / \partial x = 0$ \cite{maki1964pauli,supp}. This free-energy-based approach enables the determination of both the supercooling field and the true thermodynamic critical field in the presence of FOTs.

In the absence of ISOC ($\beta_{\mathrm{so}} = 0$), the quasiparticle spectrum under an in-plane magnetic field reduces to the conventional Zeeman-split form, $E_{\mathbf{p}\pm}=\sqrt{\Delta^2+\xi_\mathbf{p}^2}\pm H$. The corresponding phase diagram is shown in Fig.~\ref{Ising_illustration}(b). At zero temperature, the $f_s(x)$ exhibits a single minimum at $x = 1$ for $h < h_p/\sqrt{2}$, where $h_p$ is the dimensionless Pauli limit. In the intermediate field range $h_p/\sqrt{2} < h < h_p$, an additional local minimum emerges at $x = 0$, as illustrated in Fig.~\ref{Ising_illustration}(c). Notably, solving the gap equation with $\Delta \to 0$ yields the lowest field $h = h_p/\sqrt{2}$ at which $x = 0$ changes from a maximum to a local minimum of $f_s(x)$, corresponding to the supercooling field. However, the normal state becomes thermodynamically stable only when the field reaches the critical value $h = h_p$, at which the two minima at $x = 0$ and $x = 1$ are degenerate. At higher temperatures ($t = 0.40$), the $f_s(x)$ exhibits a single minimum for all $h$, as shown in Fig.~\ref{Ising_illustration}(d). In this regime, the gap equation admits only one nontrivial solution, and both the gap-equation and free-energy approaches yield the same critical field. In summary, within the FOT regime, the free-energy analysis determines the thermodynamic critical field (solid yellow curve in Fig.~\ref{Ising_illustration}(b)), while solving the gap equation at $x = 0$ only yields the supercooling field (dashed blue curve). In contrast, in the SOT regime, both methods give identical critical fields, as indicated by the solid blue curve (see the \cite{supp} for details).

\section{Thermodynamic Properties}

\begin{figure*}[ht]
\centering
\includegraphics[width=0.95\linewidth]{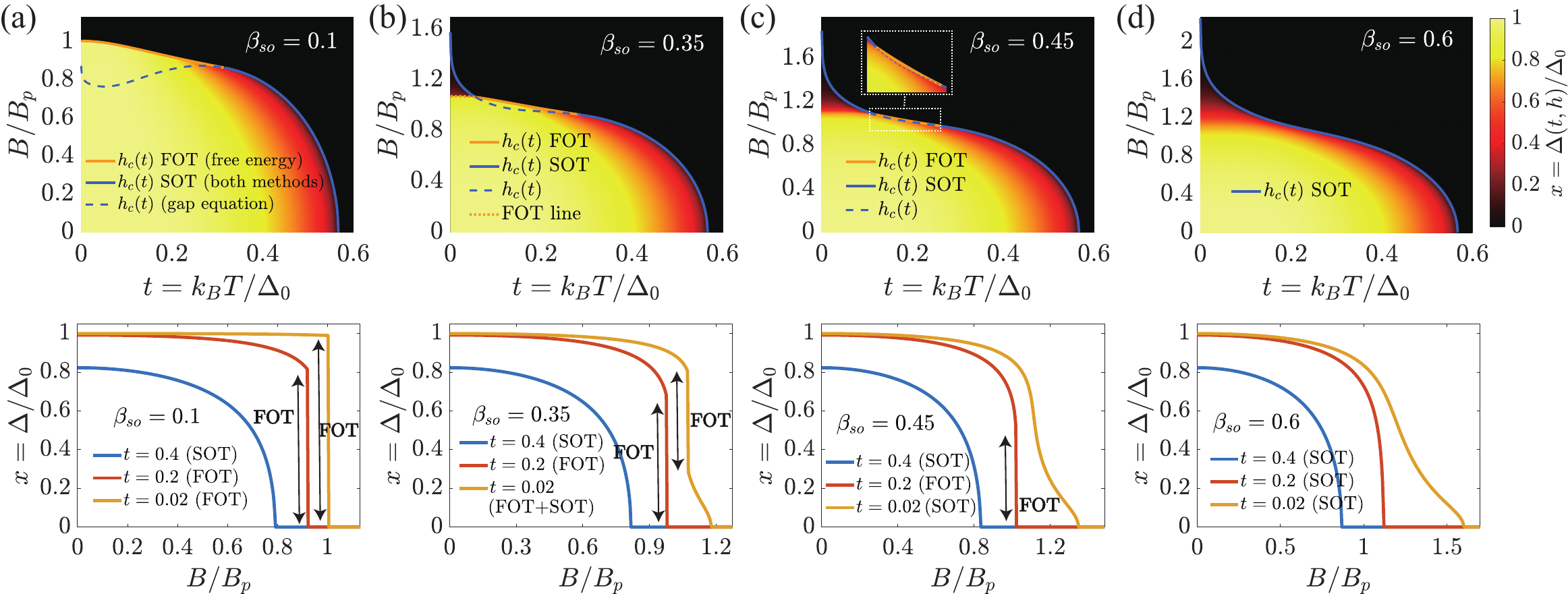}
\caption{
Field-temperature ($h$-$t$) phase diagrams and corresponding $\Delta(h)$ curves at representative temperatures ($t = 0.02, 0.2, 0.4$) for (a) $\beta_{\mathrm{so}} = 0.1$, (b) $\beta_{\mathrm{so}} = 0.35$, (c) $\beta_{\mathrm{so}} = 0.45$, and (d) $\beta_{\mathrm{so}} = 0.6$. The solid yellow curves denote the FOT critical field $h_c(t)$ obtained from free-energy analysis. The solid blue curves represent the SOT critical field. The dashed blue curves indicate the supercooling field in the FOT regime. The densely dashed silver curve in (b) marks the FOT occurring below the critical field. The color scale represents the normalized gap $\Delta(t,h)/\Delta_0$.
}
\label{h_t_phase_diagram_Ising}
\end{figure*}

Considering a finite but weak ISOC, superconductivity is not fully protected against exchange-field-induced depairing, allowing FOTs to emerge. Due to the competition between ISOC-induced protection and Zeeman depairing, the field-temperature ($h$-$t$) phase behavior exhibits distinct features that can be broadly classified into three cases, differing from both conventional superconductors and strong Ising superconductors.

For the first case with extremely weak ISOC $\beta_{\mathrm{so}} = 0.1$, the $h$-$t$ phase diagram [Fig.~\ref{h_t_phase_diagram_Ising}(a)] closely resembles that of a conventional superconductor [Fig.~\ref{Ising_illustration}(b)]. However, the supercooling fields are noticeably lifted near zero temperature, indicating that ISOC-induced protection is most effective in the low-temperature regime. The magnetic-field dependence of the superconducting gap at different temperatures [Fig.~\ref{h_t_phase_diagram_Ising}(a)] also closely follows the behavior of conventional superconductors.

Increasing the ISOC strength to $\beta_{\mathrm{so}} = 0.35$, the protective effect remains most prominent at low temperatures. As shown in Fig.~\ref{h_t_phase_diagram_Ising}(b), the critical field in the low-temperature regime (solid blue curve) slightly exceeds the Pauli limit and becomes SOTs. Unlike the conventional SOT scenario, an FOT occurs at a lower field before reaching the critical field, as indicated by the densely dashed silver curve. At intermediate temperatures, although the supercooling field approaches the critical field, the phase transition remains first order. The FOT character is further confirmed by the discontinuous jump in the gap $\Delta(h)$ at $t = 0.02$ and $0.2$. In the high-temperature regime, the transition remains second order, with a slightly enhanced critical field.

For a stronger ISOC $\beta_{\mathrm{so}} = 0.45$, the protection becomes more pronounced and extends to higher temperatures. As shown in Fig.~\ref{h_t_phase_diagram_Ising}(c), the SOT critical fields in both low- and high-temperature regimes are further enhanced. The previously observed low-temperature FOT below the critical field evolves into a smooth crossover. Consequently, the FOT region is confined to a narrow intermediate temperature range. The presence or absence of FOT is consistently reflected in the discontinuous or continuous behavior of $\Delta(h)$.

For sufficiently strong ISOC $\beta_{\mathrm{so}} = 0.6$, the protection extends across the entire temperature range, completely suppressing FOTs, as shown in Fig.~\ref{h_t_phase_diagram_Ising}(d). The absence of FOT is evidenced by the smooth evolution of $\Delta(h)$ for all temperatures. In this regime, the superconducting gap varies continuously throughout the $h$-$t$ phase space. In addition, a spectroscopic mirage gap \cite{tang2021magnetic} is expected to emerge at energies far from the Fermi level under strong in-plane magnetic fields.

\section{Spectroscopic Signatures}

The energy-dependent superconducting density of states (DOS) $N(E)$ in weak Ising superconductors can exhibit qualitatively different behavior. At zero temperature, the DOS, denoted as $N_z(E)$, can be derived from the quasiparticle dispersions in Eq.~(\ref{quasiparticle}) as
\begin{equation}\label{quasiparticle_DOS}
    N_{z}(E)=\frac{N_0}{2}\sum_{\sigma=\pm}\left|\frac{\mathrm{d}\xi_{\mathbf{p}}(E_{\mathbf{p}\sigma})}{\mathrm{d}E_{\mathbf{p}\sigma}}\right|_{E_{\mathbf{p}\sigma}=E},
\end{equation}
\noindent where $N_0$ is the normal-state DOS. A detailed derivation and the explicit analytical form of $N_z(E)$ are provided in the \cite{supp}.

\begin{figure*}[ht]
\centering
\includegraphics[width=0.9\linewidth]{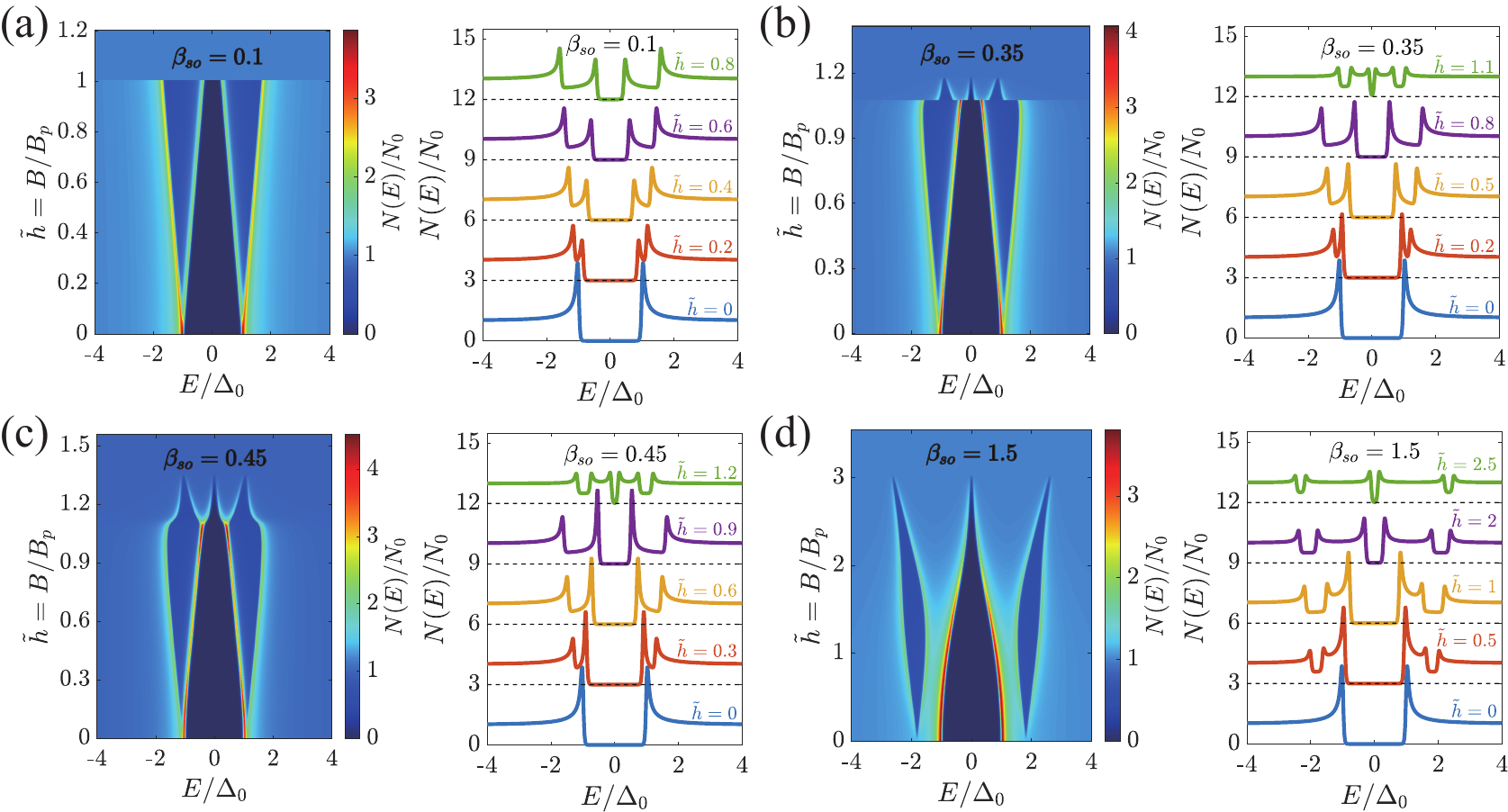}
\caption{ (a-d) Maps of DOS $N(E)$ as a function of energy and magnetic field, together with representative spectra at selected fields, for (a) $\beta_{\mathrm{so}} = 0.1$, (b) $\beta_{\mathrm{so}} = 0.35$, (c) $\beta_{\mathrm{so}} = 0.45$, and (d) $\beta_{\mathrm{so}} = 1.5$. Here, $\tilde{h} = B/B_p$ denotes the magnetic field normalized by the Pauli limit. A Dynes broadening parameter $\Gamma = 0.001\Delta_0$ and a finite temperature $t=0.02$ are used. The color scale represents $N(E)/N_0$.
}
\label{DOS_evolution}
\end{figure*}

To incorporate finite-temperature effects and inelastic scattering, we further include thermal broadening and a small Dynes parameter in the calculation of $N(E)$ \cite{dynes1978direct}. Since the ISOC-induced protection against exchange-field depairing is most effective at low temperatures and is rapidly suppressed at higher temperatures, we focus on the evolution of $N(E)$ under increasing in-plane magnetic field at $t = 0.02$ for four representative values of $\beta_{\mathrm{so}}$.

Consistent with the $h$-$t$ phase diagram resembling that of a conventional superconductor, the case $\beta_{\mathrm{so}} = 0.1$ exhibits a typical Zeeman-split superconducting DOS [Fig.~\ref{DOS_evolution}(a)], where the inner coherence peaks are weaker than the outer two. This indicates that, in the limit of extremely weak ISOC, the system effectively reduces to a conventional superconductor.

For a relatively moderate ISOC strength $\beta_{\mathrm{so}} = 0.35$, the superconducting coherence peaks remain split into four peaks at low magnetic fields [Fig.~\ref{DOS_evolution}(b)]. However, in contrast to both conventional and strong Ising superconductors, the inner two peaks become significantly sharper and more pronounced than the outer two. As discussed in the \cite{supp}, these sharp inner peaks originate from the hybridization between mirage-gap states and the inner Zeeman-split coherence peaks. This interpretation is further supported by the DOS spectrum after the FOT ($\tilde{h} = 1.1$), where six distinct peaks are observed. Therefore, the emergence of pronounced inner coherence peaks serves as a spectroscopic signature of mirage-gap states, which are associated with induced triplet pairing correlations \cite{tang2021magnetic,patil2023spectral}, in the weak-ISOC regime. Upon increasing the ISOC strength to $\beta_{\mathrm{so}} = 0.45$ [Fig.~\ref{DOS_evolution}(c)], a similar evolution of $N(E)$ is observed. However, in this case, the FOT is absent, and the in-gap mirage-gap features gradually separate from the inner coherence peaks. Notably, as discussed in the \cite{supp}, these in-gap mirage-gap features persist up to $\beta_{so}\approx 1$, where FOT has already vanished. They therefore provide a spectroscopic criterion for identifying weak Ising superconductivity over a broader parameter range than the FOT itself. For sufficiently strong ISOC $\beta_{\mathrm{so}} = 1.5$, the mirage-gap states shift outside the superconducting gap [Fig.~\ref{DOS_evolution}(d)].

\section{Superconducting Phase Diagram in the $(\beta_{\mathrm{so}},t)$ Parameter Space}

\begin{figure}[ht]
\centering
\includegraphics[width=0.8\linewidth]{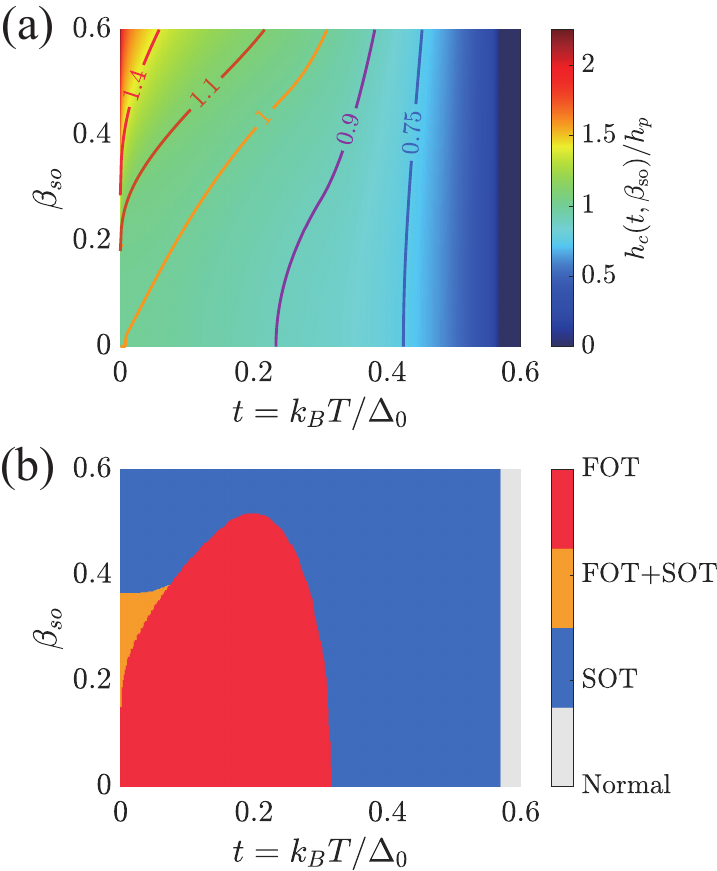}
\caption{
(a) Critical field $h_c(t,\beta_{\mathrm{so}})$ as a function of $\beta_{\mathrm{so}}$ and $t$. The color scale represents the normalized critical field $h_c(t,\beta_{\mathrm{so}})/h_p$. Contour lines highlight the enhancement of the critical field due to ISOC.
(b) Classification of superconducting behaviors in the $(\beta_{\mathrm{so}}, t)$ phase space. The red region indicates an FOT at the critical field. The orange region denotes an FOT transition occurring below a second-order critical field. The blue region corresponds to an SOT at the critical field. The white region represents the normal state.
}
\label{critical_field_types}
\end{figure}

Finally, we discuss the superconducting behavior under in-plane magnetic fields in the phase space spanned by $\beta_{\mathrm{so}}$ and $t$. We first calculate the critical field $h_c(t,\beta_{\mathrm{so}})$, as shown in Fig.~\ref{critical_field_types}(a). The contours of $h_c(t,\beta_{\mathrm{so}})$ show that increasing $\beta_{\mathrm{so}}$ enhances the critical-field enhancement predominantly at low-temperatures, whereas the enhancement becomes negligible at high temperatures.

The type of the phase transition also depends sensitively on both $\beta_{\mathrm{so}}$ and $t$. Based on the $\Delta(h)$ characteristics shown in Fig.~\ref{h_t_phase_diagram_Ising}, three representative behaviors can be identified:
(i) an FOT occurs at the critical field [e.g., $t=0.02$ in Fig.~\ref{h_t_phase_diagram_Ising}(a)];
(ii) an FOT occurs below a higher SOT-type critical field [e.g., $t=0.02$ in Fig.~\ref{h_t_phase_diagram_Ising}(b)];
(iii) an SOT occurs at the critical field [e.g., $t=0.02$ or $t=0.4$ in Fig.~\ref{h_t_phase_diagram_Ising}(c)].
Using these criteria, we classify the $(\beta_{\mathrm{so}}, t)$ phase space, as shown in Fig.~\ref{critical_field_types}(b).

For very weak ISOC ($\beta_{\mathrm{so}} \leq 0.15$), the $h$-$t$ phase diagram closely resembles that of a Pauli-limited superconductor [Figs.~\ref{Ising_illustration}(b) and \ref{h_t_phase_diagram_Ising}(a)]: an FOT occurs at the critical field in the low-temperature regime, while the transition becomes second order at higher temperatures. For relatively moderate ISOC ($0.15 < \beta_{\mathrm{so}} \leq 0.37$), the representative behavior is shown in Fig.~\ref{h_t_phase_diagram_Ising}(b): an FOT occurs below the SOT-type critical field at low temperatures, while the transition is first order (second order) in the intermediate (high) temperature regimes.

In a narrow parameter window ($0.37 < \beta_{\mathrm{so}} \leq 0.38$), ISOC provides sufficient protection at extremely low temperatures such that no FOT occurs below the second-order critical field. However, at slightly higher (yet still low) temperatures, an FOT reemerges below the second-order critical field. Owing to the extremely narrow range of $\beta_{\mathrm{so}}$ over which this behavior occurs, a representative phase diagram is not shown in Fig.~\ref{h_t_phase_diagram_Ising}.

For $0.38 < \beta_{\mathrm{so}} \leq 0.52$, the representative phase diagram is given in Fig.~\ref{h_t_phase_diagram_Ising}(c). In this regime, the transition at the critical field is second order at both low and high temperatures, while a first-order transition appears only in an intermediate temperature window. For sufficiently strong ISOC ($\beta_{\mathrm{so}} > 0.52$) \cite{ilic2017enhancement}, the FOT is completely suppressed over the entire temperature range, as shown in Fig.~\ref{h_t_phase_diagram_Ising}(d). In this case, the commonly used approach of solving the gap equation with $\Delta = 0$ \cite{yuan2016ising,liu2017unconventional,ilic2017enhancement,liu2020microscopic,mockli2020ising,wang2021ising,ilic2023spectral} correctly yields the thermodynamic critical field.

\section{Discussions and Conclusions}

To assess the experimental robustness of the predicted FOTs and spectral features, we further examine the effects of impurity scattering \cite{luttinger1960ground,virtanen2020quasiclassical,vorontsov2003thermodynamic,serene1983quasiclassical,schmitt1986transport,abrikosov1959theory,hirschfeld1988consequences,amundsen2019quasiclassical,fulde1966theory}. As detailed in the \cite{supp}, nonmagnetic intravalley and intervalley scatterings have only a weak effect on the FOT, even when the corresponding scattering rates are comparable to the superconducting gap. By contrast, sufficiently strong spin-orbit impurity scattering weakens and can ultimately eliminate the FOT. The pronounced in-gap coherence peaks are similarly robust against intravalley scattering but are progressively suppressed by intervalley and spin-orbit scattering. Nevertheless, over a substantial range of intervalley scattering and weak spin-orbit scattering, the spectra remain distinguishable from those of conventional Pauli-limited superconductors. These results indicate that the predicted thermodynamic and spectroscopic signatures survive moderate disorder, although sufficiently strong spin-orbit or other pair-breaking processes may obscure them.

In summary, we have demonstrated that FOTs can emerge in superconductors with weak ISOC within an appropriate temperature window. We further identify two pronounced in-gap coherence peaks as a spectroscopic manifestation of mirage-gap states in the weak-ISOC regime. Together, these thermodynamic and spectroscopic features distinguish weak Ising superconductors from both conventional and strong Ising superconductors. As discussed in the \cite{supp}, these distinctions gradually disappear in the limits of vanishingly weak or strong ISOC. We therefore propose several experimentally accessible criteria for identifying the weak-ISOC regime. A modest enhancement of the low-temperature critical field above the Pauli-limited value, for example by approximately $10\%$ within the present model \cite{supp}, provides a practical initial indication. More distinctive signatures are: (i) an FOT confined to an intermediate temperature range, with SOTs at both lower and higher temperatures, and (ii) pronounced inner coherence peaks in the quasiparticle spectrum. The former may be detected through magnetic-field hysteresis \cite{tedrow1977supercooling} or thermodynamic anomalies associated with a discontinuous transition \cite{bianchi2002first}, whereas the latter can be directly probed by tunneling spectroscopy or scanning tunneling microscopy. Consequently, the simultaneous observation of pronounced in-gap inner coherence peaks, together with a moderate critical-field enhancement and an intermediate-temperature FOT regime, provides a combined criterion for identifying weak Ising superconductivity.

The weak ISOC may be realized through several experimental routes. Increasing the layer number in ISOC superconducting films can reduce the effective ISOC strength \cite{de2018tuning}, while proximity to transition-metal dichalcogenides can induce ISOC in otherwise conventional superconductors \cite{wu2019induced}. In addition, several layered or transition-metal-dichalcogenide-based superconductors exhibit only modest enhancements of the in-plane critical field above the Pauli limit \cite{ji2024continuous,qu2025strongly}. These systems may therefore provide candidate platforms for exploring the weak-ISOC regime. 

\begin{acknowledgements}
X. Wang acknowledges X.-Y. Zhou, Z.-C. Zhang, Y.-Q. Yan, J.-B. Liu, and M. Shu for stimulating discussions. This work is financially supported by Quantum Science and Technology-National Science and Technology Major Project (Grants No. 2023ZD0300500) and the National Natural Science Foundation of China (Grants No. 11427903 and 52388201).\\
\end{acknowledgements}

\nocite{*}

\bibliographystyle{cite_form}
\bibliography{bib_main}

\end{document}